\documentclass[11pt,a4paper,extrafontsizes,oneside]{article}
\pdfoutput=1
\usepackage[scale=0.7,vmarginratio={1:2},heightrounded]{geometry}
\usepackage{caption}
\DeclareCaptionType{copyrightbox}
\usepackage{multirow}
\usepackage{float}
\usepackage{array}
\newcolumntype{L}{>{\centering\arraybackslash}m{2cm}}
\usepackage{subcaption}
\usepackage{url}
\usepackage{graphicx}
\usepackage{epstopdf}
\graphicspath{{./figures/}}
\usepackage[usenames,dvipsnames]{color}
\usepackage[utf8]{inputenc}
\usepackage[table]{xcolor}\usepackage[symbol]{footmisc}
\usepackage{algorithm}
\usepackage[noend]{algpseudocode}
\DeclareUnicodeCharacter{00A0}{ }

\begin{document}

\linespread{0.95}
\setlength{\parskip}{0pt}
\setlength{\parsep}{0pt}
\setlength{\topskip}{1pt}
\setlength{\topsep}{0pt}
\setlength{\partopsep}{0pt}

\makeatletter
\let\@copyrightspace\relax
\makeatother

\title{EMBERS at 4 years:\\
Experiences operating an \\
Open Source Indicators Forecasting System}

\author{
   Sathappan Muthiah\footnotemark[1],
  Patrick Butler\footnotemark[1],
  Rupinder Paul Khandpur\footnotemark[1],\\
  Parang Saraf\footnotemark[1],
  Nathan Self\footnotemark[1],
Alla Rozovskaya\footnotemark[1],
  Liang Zhao\footnotemark[1],
  Jose Cadena\footnotemark[2],\\
  Chang-Tien Lu\footnotemark[1],
  Anil Vullikanti\footnotemark[2],
  Achla Marathe\footnotemark[2],
  Kristen Summers\footnotemark[3],
  Graham Katz\footnotemark[4], \\
  Andy Doyle\footnotemark[4],
  Jaime Arredondo\footnotemark[5],
  Dipak K. Gupta\footnotemark[6],
  David Mares\footnotemark[5],
  Naren Ramakrishnan\footnotemark[1],
\\
}

\maketitle

\footnotetext[1]{Discovery Analytics Center, Virginia Tech, Arlington,VA 22203}
\footnotetext[2]{Biocomplexity Institute, Virginia Tech, Blacksburg, VA 24061}
\footnotetext[5]{University of California at San Diego, San Diego, CA 92093}
\footnotetext[6]{San Diego State University, San Diego, CA 92182}
\footnotetext[4]{CACI Inc., Lanham, MD 20706}
\footnotetext[3]{IBM Watson Group, Chantilly, VA 20151}

\begin{abstract}
EMBERS is an anticipatory intelligence system forecasting population-level events in multiple
countries of Latin America. A deployed system from 2012, EMBERS has been generating alerts
24x7 by ingesting a broad range of data sources including news, blogs, tweets, machine coded events,
currency rates, and food prices.
In this paper, we describe our experiences operating EMBERS continuously for nearly 4 years,
with specific attention to the discoveries it has enabled, correct as well as missed
forecasts, and lessons learnt from participating in a forecasting tournament including
our perspectives on the limits of forecasting and ethical considerations.
\end{abstract}

\section{Introduction}
Modern communication forms such as social media and microblogs are not only rapidly
advancing our understanding of the world but also improving the methods by which we can
comprehend, and even forecast, the progression of events.
Tracking population-level activities via `massive passive' data has been shown to
quite accurately shed light into large-scale societal movements.

Two years ago, in KDD 2014, we described EMBERS~\cite{kdd:beating-the-news}, a deployed anticipatory
intelligence system~\cite{bigdata-andy-doyle-embers-paper} that forecasts significant
societal events (e.g., civil unrest
events such as protests, strikes, and `occupy' events) using a large set of open source
indicators such as news, blogs, tweets, food prices, currency rates, and other public
data. The EMBERS system has been running continuously 24x7 for nearly 4 years at this point
and our goal in this paper is to present the discoveries it has enabled,
both correct as well as missed
forecasts, and lessons learned from participating in a forecasting tournament including
our perspectives on the limits of forecasting and ethical considerations. In
particular, we shed insight into the value proposition to an analyst and how EMBERS forecasts
are communicated to its end-users.

The development of EMBERS is supported by the Intelligence Advanced Research Projects
Activity (IARPA) Open Source Indicators (OSI) program.
EMBERS forecasts are scored against the Gold Standard Report (GSR), a monthly catalog of
events as reported in newspapers of record in 10 Latin American
countries - Argentina, Brazil, Chile, Colombia, Ecuador, El Salvador,
Mexico, Paraguay, Uruguay, Venezuela. The GSR is compiled by MITRE corporation
using human analysts.
EMBERS currently focuses on multiple regions of the world but for the purpose of this paper
we focus primarily on Latin America, specifically the countries of
Argentina, Brazil, Chile, Colombia, Ecuador, El Salvador, Mexico, Paraguay, Uruguay, and Venezuela.
Similarly, EMBERS generates forecasts for multiple event classes---influenza like illnesses~\cite{prithwish-ili},
rare diseases~\cite{sdm-saurav}, elections~\cite{aravindan-wei-besc}, domestic political crises~\cite{gdelt-acm-webscience}, and civil unrest---but in this paper we focus primarily on civil unrest as this was the
most challenging event class with hundreds of events every month across the countries studied here.

Our key contributions can be summarized as follows:
\begin{enumerate}
\setlength\itemsep{0pt}
\itemsep0em
\item   Unlike retrospective studies of predictability, EMBERS forecasts are communicated in real-time before the
event to MITRE/IARPA and scored independently of the authors.
We present multiple quantitative indicators of EMBERS performance
as well as insights into how
we made EMBERS forecasts most valuable to analysts. We report two primary
ways in which analysts utilize EMBERS and the use of {\it automated
narratives} to help make EMBERS forecasts as useful as possible.

\item
In an attempt to demystify the state-of-the-art
in forecasting and to create an open dialogue in the community, we report both successful
forecasts of EMBERS as well as events missed by EMBERS. The events not forecast by EMBERS lead us to
considerations of both the limitations of the underlying technology as well as
the inherent limits to forecasting large-scale events.
\item
While social media is often touted as the key to event forecasting
systems such as EMBERS, we present the results of an ablation study to outline
the performance degradation that ensues if data sources
such as Twitter and Facebook were to be removed from the forecasting pipeline.
\item
We consider the separation of civil unrest events into events
that happen with a degree of regularity versus rare
or significant
events, and evaluate the performance of EMBERS in forecasting such
surprising events.
\item
We describe our current best understanding of the limitations to
forecasting civil unrest events using technologies like EMBERS and
also consider the ethical considerations of the EMBERS technologies.
\end{enumerate}

\section{Background}
\label{sec:background}
We begin by providing a brief review of forecasting systems, followed by a
quick preview of EMBERS, its system architecture, machine learning models,
and measures for evaluating its performance. For more details, please
see~\cite{kdd:beating-the-news}.

Forecasting societal events such as civil unrest has a long tradition in the intelligence analysis and political science
community. We distinguish between forecasting systems versus event coding
systems (systems that provide
structured representations of ongoing events reported in newspapers), and focus on the former.
Early forecasting systems such as ICEWS~\cite{icews} provided very broad coverage in countries but
were limited by their spatio-temporal resolution (e.g., typically country- and month- level forecasting for
specific events of interest~\cite{eoiprediction}). The ICEWS events of interest are
domestic political crises, international crises, ethnic/religious violence, insurgencies,
and rebellion.
A similar project in scope is PITF (Political Instability
Task Force)~\cite{pitf} funded by the CIA.
To the best of our knowledge, only EMBERS provides
the most specific spatial resolution (city-level) and temporal resolution (daily-level) capability in forecasting.

The software architecture of EMBERS (Early Model Based Event Recognition using
Surrogates) is designed as a loosely coupled, share-nothing, highly distributed pipeline of
processes connected via ZeroMQ.  In this manner, the system is both highly scalable and fault
tolerant.  The EMBERS pipeline can loosely be broken up into four stages:
ingestion, enrichment, modeling, and selection.  In the first stage, ingestion,
data is collected from a variety of sources and streamed into the following
stages in real-time.  The enrichment stage takes the raw data from the ingestion stage
and processes it in various ways including natural language processing,
geocoding, and relative time phrase normalization.  After enrichment, the
modeling stage feeds the enriched data into the various models that make up
EMBERS.  Unlike other systems which use single monolithic models to make
predictions, EMBERS combines the results of several different models to arrive
at the most accurate forecasts.  Finally, in the selection stage the separate
alerts from each model are de-duplicated, fused, and selected and finally
emitted as a full forecast for a real world event.

\begin{figure}
\centering
  \includegraphics[width=0.7\columnwidth]{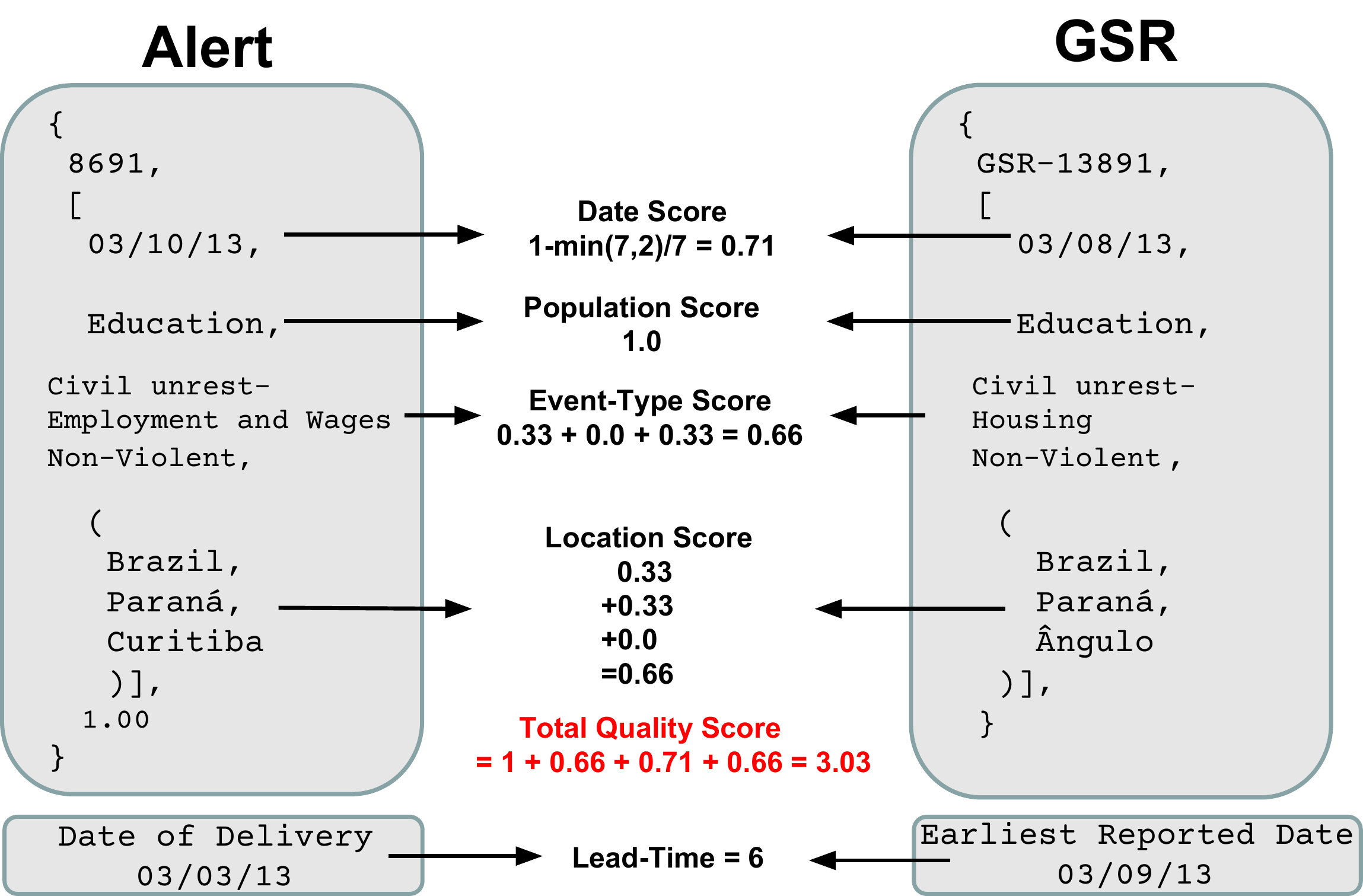}
\caption{An example depicting how an alert is scored with respect to the ground truth.}
\label{fig:alert}
\end{figure}

\begin{figure}
\includegraphics[width=\columnwidth]{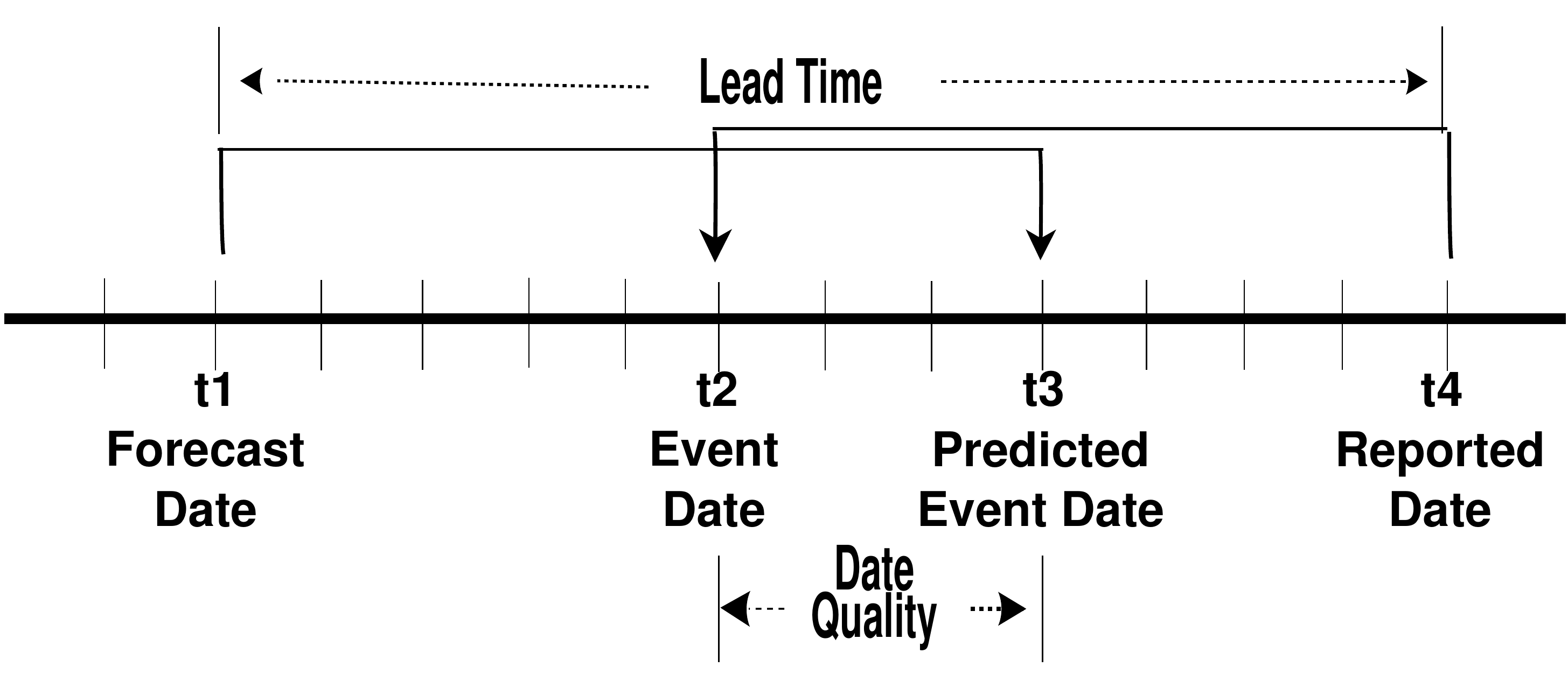}
\caption{Alert sent at time $t1$ predicting an event at time $t3$
can be matched to a GSR event that happened at time $t2$ and reported
at time $t4$ if $t1 < t4$.}
\label{fig:leadtime}
\end{figure}

The structure of a civil unrest forecast is shown in
Figure~\ref{fig:alert} (left).
A forecast constitutes four fields, corresponding to the when, where, who, and why
of the protest. These fields are respectively denoted as the date, location, population, and event type.
Location is recorded at the city level. Population and event type are fields chosen from a categorical
set of possibilities.  The figure further shows how an alert with all
these fields are scored against a GSR event. In the basic scoring
methodology shown in Figure~\ref{fig:alert} each of the four fields
are weighted uniformly and a total quality score out of 4 is
obtained.  Apart from this each alert also has a lead-time associated with
it calculated as shown in Figure~\ref{fig:leadtime}.

Rather than design one model to integrate all possible data sources, EMBERS adopted a multi-model
approach to forecasting. Each model utilized a specific (possibly overlapping) set of data sources
and is tuned for high precision, so that the union of these models can be tuned for high recall.
A fusion/suppression engine~\cite{andy-scotland-paper} allows a tunable
strategy to issue more or fewer
alerts depending on whether the analyst's objective is to obtain a higher
precision or recall. The underlying models used in EMBERS are: (i)
\textit{planned protest model}~\cite{pp-paper1},
(ii) \textit{dynamic query expansion}~\cite{dqe-plosone}, (iii) \textit{volume-based model}~\cite{asonam},
(iv) \textit{cascade regression}~\cite{anil-plosone}, and (v) a baseline model. The planned protest model,
for news and social media (Twitter, Facebook), identifies explicit signs of organization and calls
for protest, resolves relative mentions of time (e.g., `next Saturday') and space (e.g., `the square')
to issue forecasts. The dynamic query expansion (DQE) model uses Twitter as a data source and learns time- and country-specific
expansions of a seed set of keywords to identify specific situational circumstances for civil unrest.
For instance, in Venezuela (an economy where the government exercises stringent price controls),
there were a series of protests in 2014 stemming from the shortage of toilet paper, a novel circumstance
that was uncovered by DQE. The volume-based model uses a range of data sources, spanning
social, economic
and political indicators. It uses classical statistical models (LASSO
and hybrid regression models) to forecast civil unrest events using features
from social media (Twitter and blogs), news sources,
political event databases (ICEWS and GDELT~\cite{gdelt}), Tor~\cite{tor} statistics, food prices, and currency
exchange rates. It aims to provide a multi-source perspective into forecasting by leveraging
the selective superiorities of different data sources.  The cascade regression model
aims to model activity related to organization and mobilization in Twitter~\cite{anil-plosone}.
Finally, the baseline model uses a maximum likelihood
estimation over the GSR to issue history-based forecasts.

The EMBERS project is unique not just in its algorithmic underpinnings but also in the use of new measures
for evaluation, specifically aimed at determining forecasting performance. As
shown in Figure~\ref{fig:leadtime},
one of the primary measures of EMBERS performance is lead time, the number of days by which a forecast
`beats the news', i.e., the date of reporting of the event. Lead time should not be confused with
date quality, i.e., the difference between the predicted date and the actual date of the event. The date
quality is one of the components to the quality score, the other components being the
location score, event type score, and population score.
Figure~\ref{fig:alert} 
shows how these other
components are scored between an EMBERS forecast and a GSR record. Given a set of alerts and a set of
GSR events for a given month, the lead time is used as a constraint to define legal (alert, event) pairs so that
we can construct a bipartite matching to optimize the best quality score. From this bipartite matching,
measures of precision and recall can be derived, i.e., by assessing the number of (un)matched events or
alerts. Finally, a confidence score is used to assess the quality of probabilities imputed by EMBERS to its
forecasts, and measured in terms of the Brier score. For more details, please
see~\cite{kdd:beating-the-news}.

We now turn to a discussion of specific discoveries enabled by EMBERS, into civil unrest in Latin America, and into
the complexity of the forecasting enterprise as a whole.

\section{Performance Analysis}
\label{sec:perf}
First, we begin with a performance analysis of EMBERS, from both a quantitative
point of view with respect to the GSR and with respect to end-user (analyst) goals.
\subsection{Quantitative Metrics}
\begin{figure}

\definecolor{LightCyan}{rgb}{0.85,.95,.95}
\definecolor{yes}{RGB}{26,175,84}
\definecolor{no}{RGB}{253,191,45}
\definecolor{close}{RGB}{148,206,88}

\resizebox{\columnwidth}{!}{
\rowcolors{5}{gray!20}{white}
\begin{tabular}{|l|c|c|c|}
    \multicolumn{4}{c}{\textbf{Targets}} \\
    \hline \rowcolor{LightCyan}
                           & \textbf{Month 12}  & \textbf{Month 24} & \textbf{Month 36} \\
                                        \hline
    Mean Lead-Time         & 1 day    & 3 days   & 7 days \\
    \hline
    Mean Probability Score & 0.60     & 0.70     & 0.85 \\
    \hline
    Mean Quality Score     & 3.0      & 3.25     & 3.5 \\
    \hline
    Recall                 & 0.50     & 0.65     & 0.80 \\
    \hline
    Precision              & 0.50     & 0. 65    & 0.80 \\
    \hline
\end{tabular}
}
\resizebox{\columnwidth}{!}{
\rowcolors{2}{gray!20}{white}
\begin{tabular}{|l|c|c|c|}
    \multicolumn{4}{c}{\textbf{Actual}} \\
    \hline \rowcolor{LightCyan}
    \textbf{Metric}                 & \textbf{Month 12}  & \textbf{Month 24} & \textbf{Month 36} \\
    \hline
    Mean Lead-Time         & \cellcolor{yes} 3.89 days & \cellcolor{yes} 7.54 days & \cellcolor{yes} 9.76 days \\
    \hline
    Mean Probability Score & \cellcolor{yes} 0.72      & \cellcolor{yes} 0.89      & \cellcolor{yes} 0.88 \\
    \hline
    Mean Quality Score     & \cellcolor{no} 2.57       & \cellcolor{no}3.1         & \cellcolor{close}3.4 \\
    \hline
    Recall                 & \cellcolor{yes} 0.80      & \cellcolor{close} 0.65    & \cellcolor{yes} 0.79 \\
    \hline
    Precision              & \cellcolor{yes} 0.59      & \cellcolor{yes} 0.94      & \cellcolor{yes} 0.87 \\
    \hline
\end{tabular}
}
\caption{IARPA OSI targets and results achieved by EMBERS}
\label{fig:quant}
\end{figure}
Figure~\ref{fig:quant} depicts both the targets set by the IARPA OSI program as well as the
actual measures achieved by the EMBERS system. As shown here, the easiest target to achieve
in EMBERS was, surprisingly, the lead time objective. This was feasible due to EMBERS's focus on modeling
both planned and spontaneous events. Planned events are sometimes organized with as many as several weeks
of lead time and thus identifying indicators of organization was instrumental in achieving
lead time objectives. The confidence (mean probability) scores were also achieved by EMBERS and involved
careful calibration of probabilities by taking into account estimates of
model propensities and data source reliabilities. The measure that was most difficult to achieve
was the quality score as it involved a four component additive score and thus tangible improvements in
score required more than incremental improvements in forecasting specific components. Finally, recall
and precision involve a natural underlying trade-off and the deployment of our fusion/suppression
engine provided the ability to balance this trade-off to meet IARPA OSI's objectives.

\begin{figure}[t]
\centering
\includegraphics[width=.8\columnwidth]{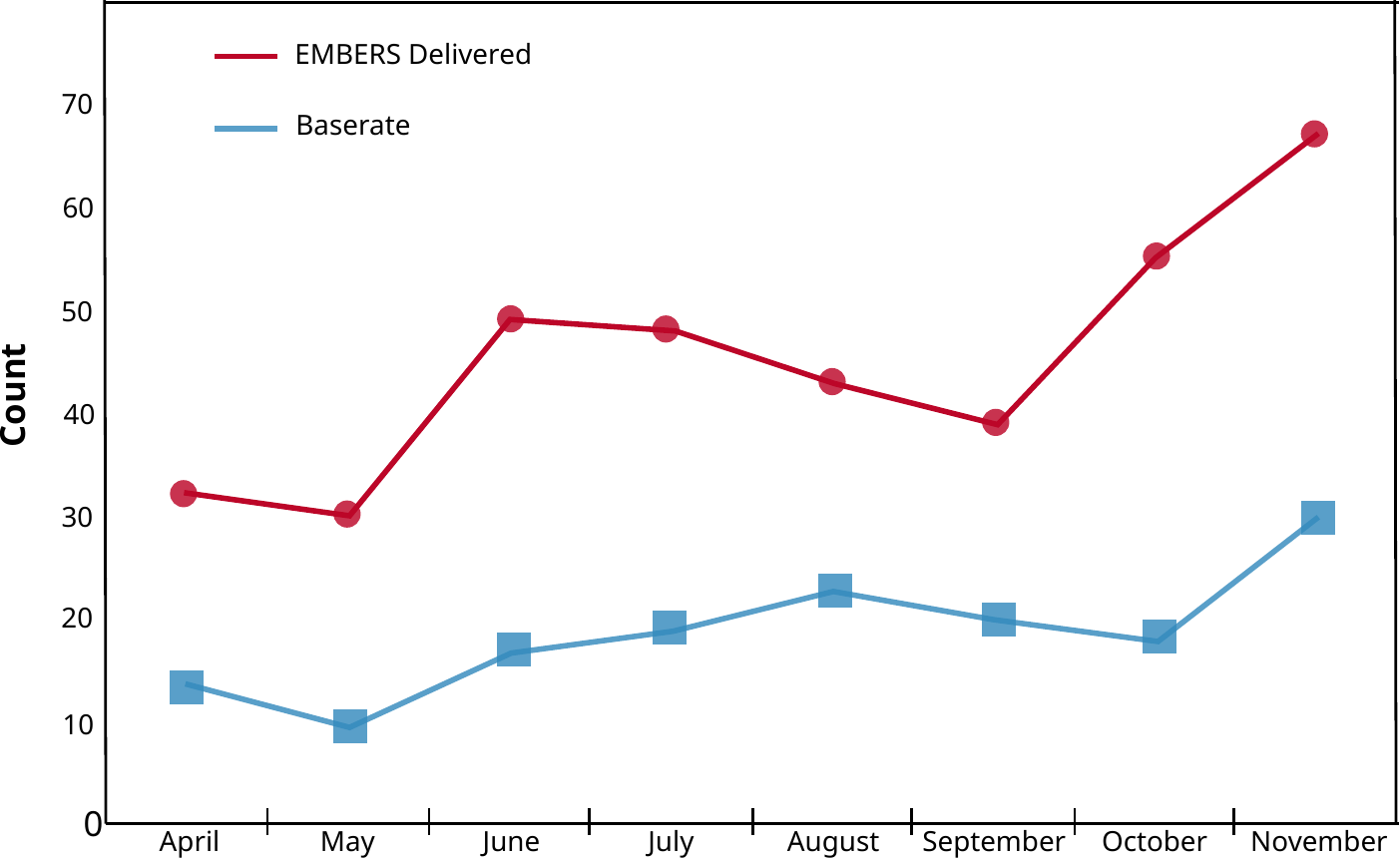}
\caption{Comparison of number of perfect scores (4.0) obtained by EMBERS vs
a baserate model each month in 2013.}
\label{fig:perfect_score}
\end{figure}

Apart from comparing mean scores another interesting metric is to see
how many perfect matches (4.0 quality score) are obtained by an
algorithm.Figure~\ref{fig:perfect_score} shows the number of alerts issued
by EMBERS that matched perfectly to an event in the future on a monthly
basis for 2013.  The figure clearly shows that EMBERS makes almost double
the number of fully accurate forecasts as compared to the baserate
model.

\subsection{Analyst Evaluation}
In addition to the quantitative measures above, our experience interacting with analysts (across multiple
branches of government) demonstrated
an interesting dichotomy as to how analysts use EMBERS alerts. Some analysts preferred to use EMBERS in an
`analytic triage' scenario wherein they could tune EMBERS for high recall so that they would apply their
traditional measures of filtering and analysis to hone in on forecasts of interest. Other analysts
instead viewed EMBERS as a data source and preferred to use it in a high precision mode, e.g., wherein they
were focused on a specific region of the world (e.g., Venezuela) and aimed to investigate a particular
social science hypothesis (e.g., whether disruptions in global oil markets led to civil unrest).

\begin{figure}[h]
\centering
\includegraphics[height=.35\textheight]{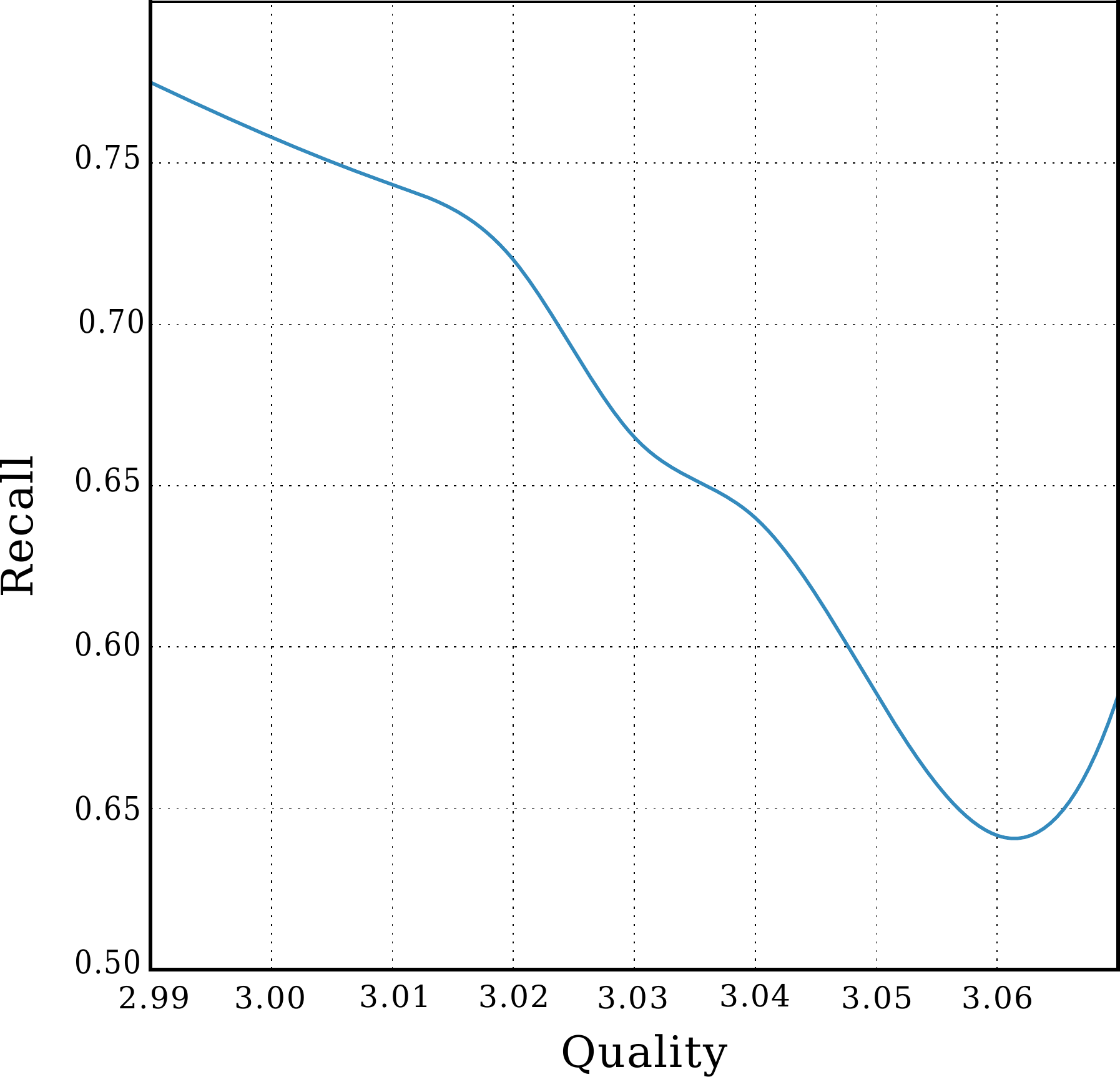}
\caption{Recall vs quality tradeoff in EMBERS.}
\label{fig:recallVsQS}
\end{figure}

To support these diverse classes of users, we implemented two
mechanisms in the alert delivery stage. First, we implemented a mechanism wherein in addition to generating alerts, EMBERS
also forecasted the expected quality score for each forecast (using machine learning methods trained on
past GSR-alert matches). This expected quality score measure provided a way for analysts to use quality
directly as a way to tune the system to receive greater or fewer
alerts.  Figure~\ref{fig:recallVsQS} shows the trade-off between final quality
score and recall when alerts are suppressed based on expected quality.
As expected we can see that the recall drops and quality increases as
the cut-off threshold for expected quality is increased.

\begin{figure}     \centering
    \fbox{
\begin{minipage}{\columnwidth}{
EMBERS forecasts that there will be a {\color{red}violent} protest on
{\color{red}February, 18th 2014}
in {\color{red}Caracas}, the {\color{red}capital city of Venezuela}.
We predict that the protest will involve
people working in the {\color{red}business sector}. The protest will be related to
{\color{red}discontent about economic policies}.
\\
There were {\color{blue}5, 5, and 5 other similar
warnings in last 2, 7 and 30 days}, respectively.
\\
The forecast date of the warning falls in {\color{blue}week 7}, which
{\color{blue}may have historical
importance}; this {\color{blue}week is found to be statistically significant}
(pval=0.00461919415894, zscore=2.832, avg. count=57.25, mean=21.569 +/- 12.597)
\\
Audit trail of the warning includes an {\color{blue}article printed 2014-02-17}.
\\
\underline{Major players} involved in the protest include {\color{red}Venezuelan opposition leader, students,
President Nicolas Maduro, and Leopoldo Lopez}.
\\
\underline{Reasons}: Protest {\color{magenta}against rising
inflation and crime}; Protestors want a {\color{magenta}political change}; President Nicolas
Maduro has {\color{magenta} accused US consular officials} and {\color{magenta}right-wing}.
\\
\underline{Protests are characterized by}: Venezuelan opposition leader
spearheaded days of protest and {\color{green}calling for peaceful
demonstration}; Maduro accused official on {\color{red} 2014-12-16}; Protests
have seen {\color{green}several deadly street protests}; Three people were
{\color{green}killed on} {\color{red}2014-02-12}; {\color{green}Demonstrations}
setting days of clashes; {\color{green}supporters to march to}
{\color{red}Interior Ministry} {\color{green}on} {\color{red}2014-02-18}.
}
\end{minipage}
}
    \caption{An example narrative for a EMBERS alert message. Here, color
{\color{red}red} indicates named entities, {\color{green} green} refers to
descriptive protest related keywords. Items in {\color{blue} blue} are historical or
real time statistics and those in {\color{magenta}magenta} refer
to inferred reasons of protest. }
    \label{fig:narrative}
\vspace{-5mm}
\end{figure}
Second, we implemented an automated narrative generation capability
(see Fig.~\ref{fig:narrative}) wherein EMBERS auto-generates a summary
of the alert in English prose. As shown in Fig.~\ref{fig:narrative},
a narrative comprises many parts drawn from different sources
of information. One source is the named entities and the system uses
`Wikification' to identify definitions and descriptions of these
named entities on Wikipedia. A second source is historical (or real-time)
statistics of warning output and warning performance and situating the
alert in this context. The third source pertains to inferred reasons for
the protest using knowledge graph identification techniques.

\section{EMBERS Successes and Misses}
\label{sec:success}
Next, we detail
some of the successful as well as not so successful forecasts made by EMBERS over
the past few years in Latin America.
\subsection{Successful Forecasts}
\subsubsection*{Brazilian Spring (June 2013)}
These protests were the largest and most
significant protests in Brazil's recent history and caught worldwide
attention. Millions of Brazilians took part in these demonstrations,
also known as the Brazilian Spring or the Vinegar Movement (inspired
from the use of vinegar soaked cloth by demonstrators to protect
themselves from police teargas). These protests were
sparked by an increase in public transport fares from $R\$3$
to $R\$3.20$ by the government of President Dilma Rousseff.

\begin{figure} \centering
\includegraphics[width=.8\columnwidth]{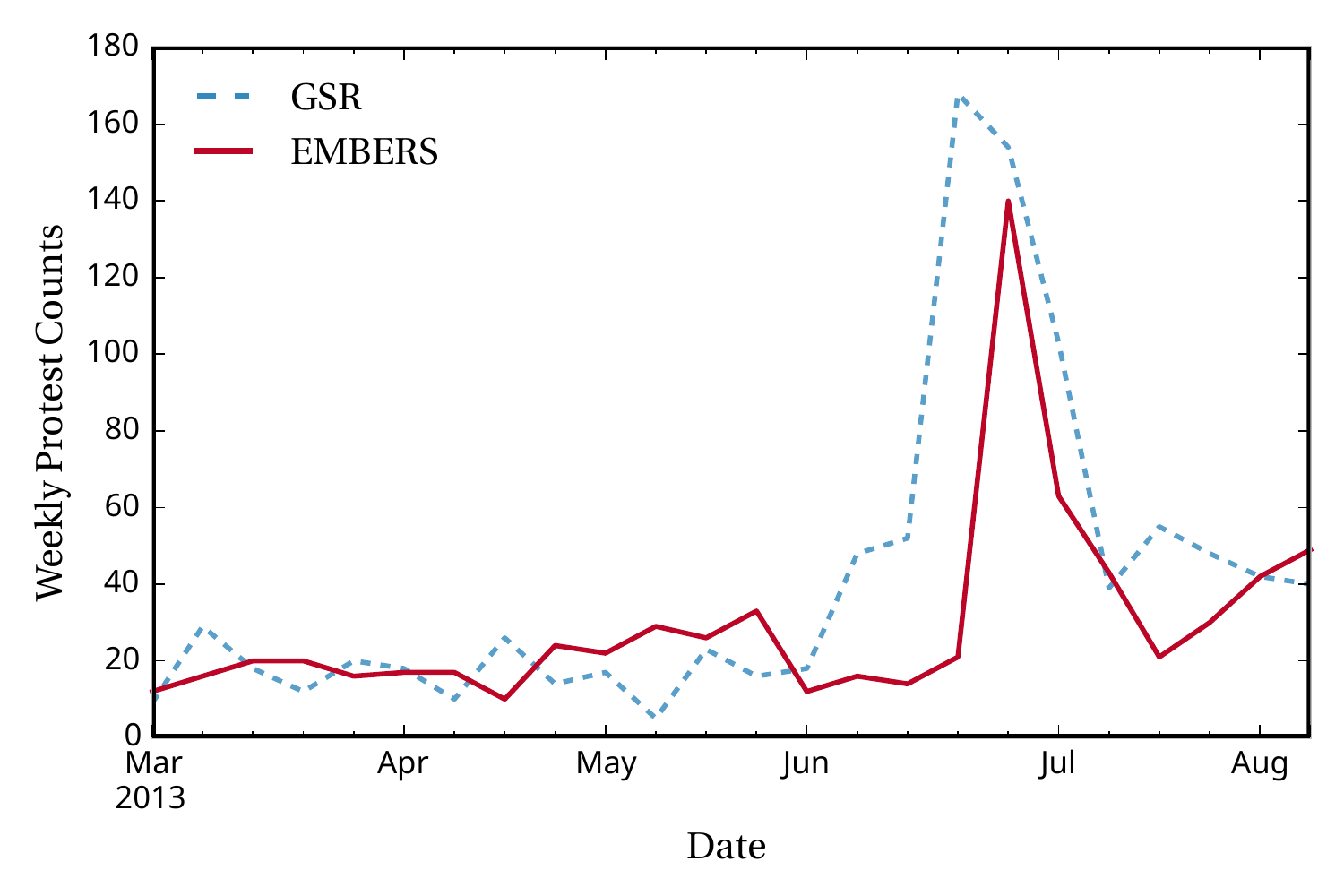}
\caption{EMBERS performance during the Brazilian Spring (June 2013.)}
\label{fig:brazilJune13}
\end{figure}

As shown in Figure~\ref{fig:brazilJune13} EMBERS, while missing the initial uptick,
forecast the increase in the order-of-magnitude of protest events
during the Brazilian Spring and also captured
the spatial spread in the events, in addition to forecasting that this event
will span the broad Brazilian general population (as opposed to being confined to
specific sectors).

\begin{figure} \centering
\includegraphics[width=.8\columnwidth]{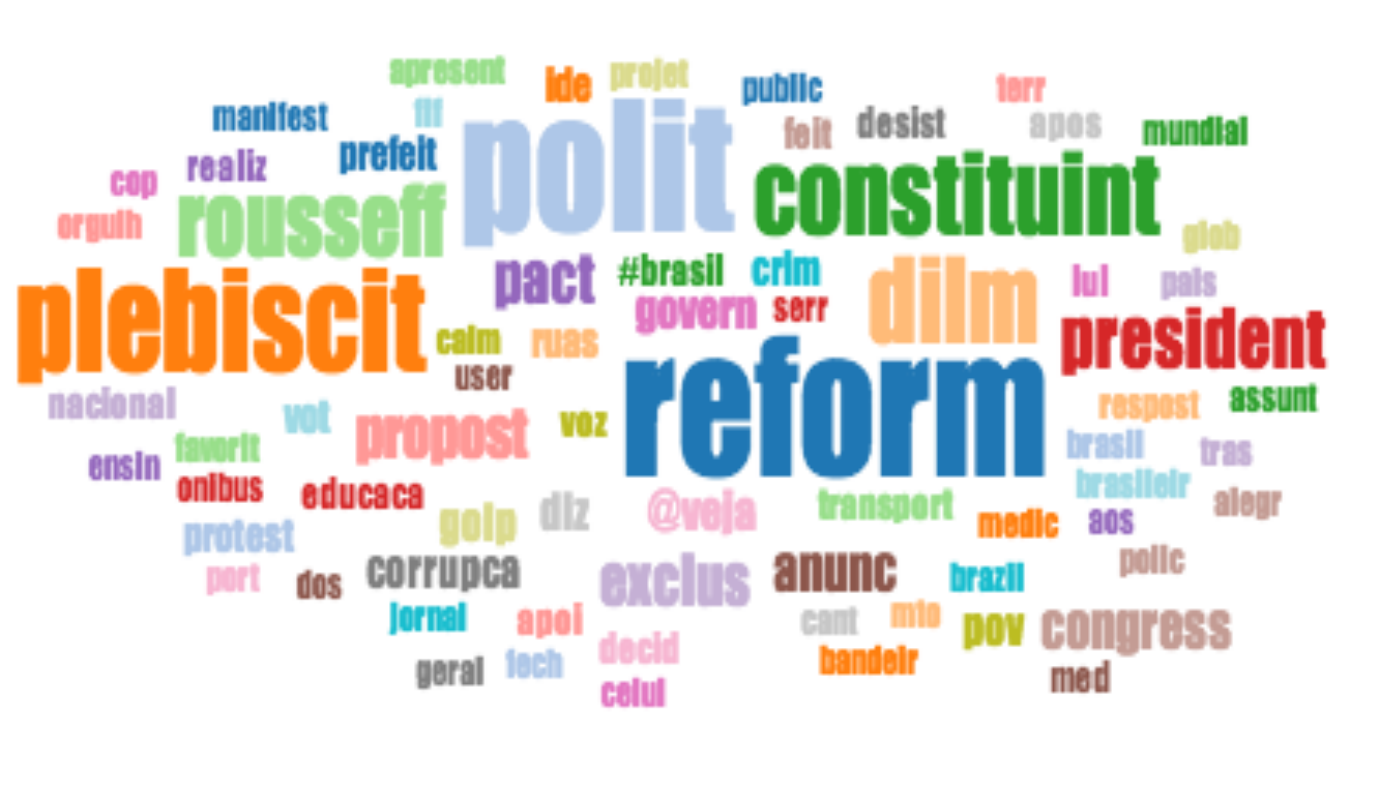}
\caption{Word cloud representing tweets identified by EMBERS dynamic query expansion model.}
\label{fig:brazilJune13_wordCloud}
\end{figure}

Around 68\% of EMBERS alerts during this period originated from
the planned protest model.
This is due to the fact that
social networking
platforms (Twitter and Facebook) as well as conventional news media played a key
role in organization of these uprisings. Although initial protests were
primarily due to the bus fare increases, they quickly morphed into
more broader dissatisfaction to include wider issues such as
government corruption, over-spending, and police brutality. The
demonstrators also made calls for political reforms. In response, President
Rousseff proposed a plebiscite on widespread political reforms in
Brazil (but this was was later abandoned). Through its dynamic query expansion model,
EMBERS was able to capture such discussions on Twitter
(see Figure~\ref{fig:brazilJune13_wordCloud}), and tracked their evolution as events
unfolded through June.

\begin{figure} \centering
\includegraphics[height=.6\columnwidth]{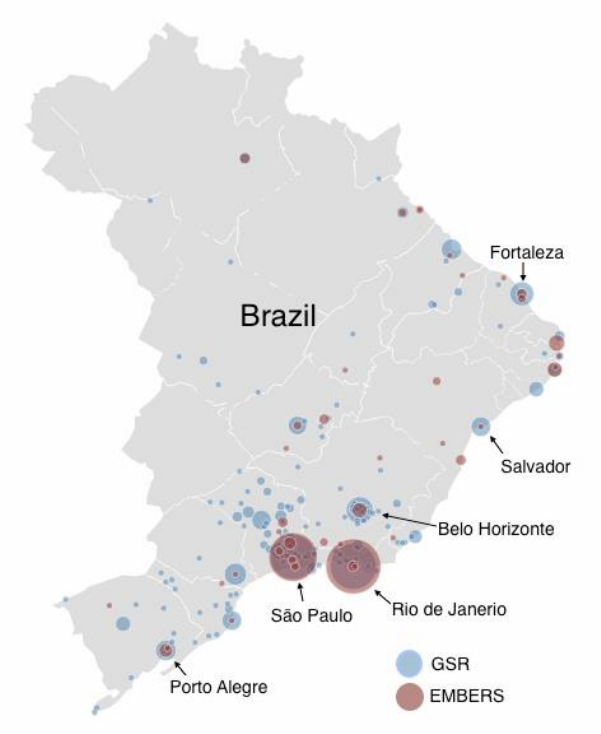}
\caption{Geographic overlap of protest events (from the GSR) and EMBERS
alerts for Brazil during June 2013.}
\label{fig:brazilJune13_map}
\end{figure}

The protests intensified in late June (see Figure~\ref{fig:brazilJune13}), which were
forecast correctly by EMBERS, and these events
also coincided with FIFA 2013 Confederation Cup matches. We believe this was an important factor in helping
the protests gain momentum,
as the events were covered by international media. A majority of protests
occurred in the cities that were hosting FIFA soccer matches.
EMBERS issued most of its alerts for these host cities (see
Figure~\ref{fig:brazilJune13_map}), viz.
Rio de Janeiro, São Paulo,
Belo Horizonte, Salvador, and Porto Alegre,
among others. For example, on 27th June during the Confederations Cup
semi-final in Fortaleza, around 5000 protesters clashed with the police
near the Castelao stadium. In this case EMBERS had forecast an alert the
day before. Later on the 30th of June, when the last games of the
confederation cup took place in Rio de Janeiro and Salvador, they
were plagued by mass protests as well; EMBERS predicted these events and
submitted multiple alerts for Rio for the 28th and 29th June and one for
Salvador for 29th June.

\subsubsection*{Venezuelan Protests (Feb-March 2014)}
\begin{figure} \centering
\includegraphics[width=.6\columnwidth]{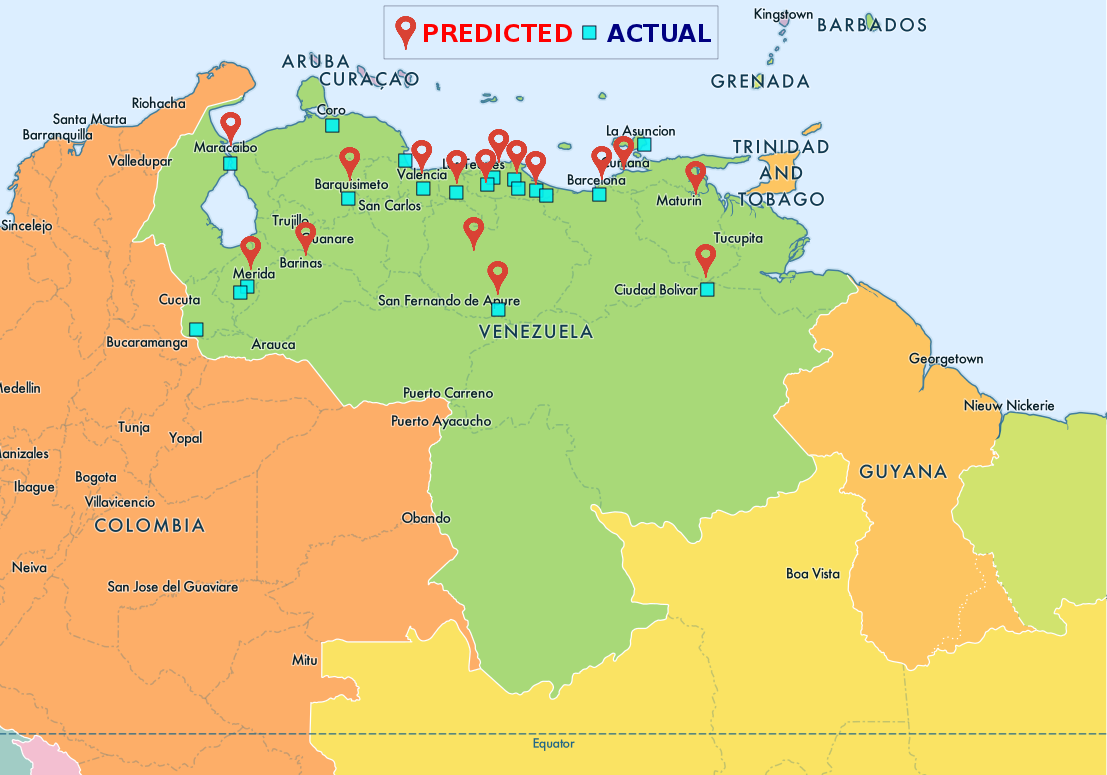}
\caption{Geographical spread of protests (and forecasts) during
Venezuelan student protests (Feb-Mar 2014).}
\label{fig:venezuelaMap}
\end{figure}

\begin{figure} \centering
\includegraphics[width=.8\columnwidth]{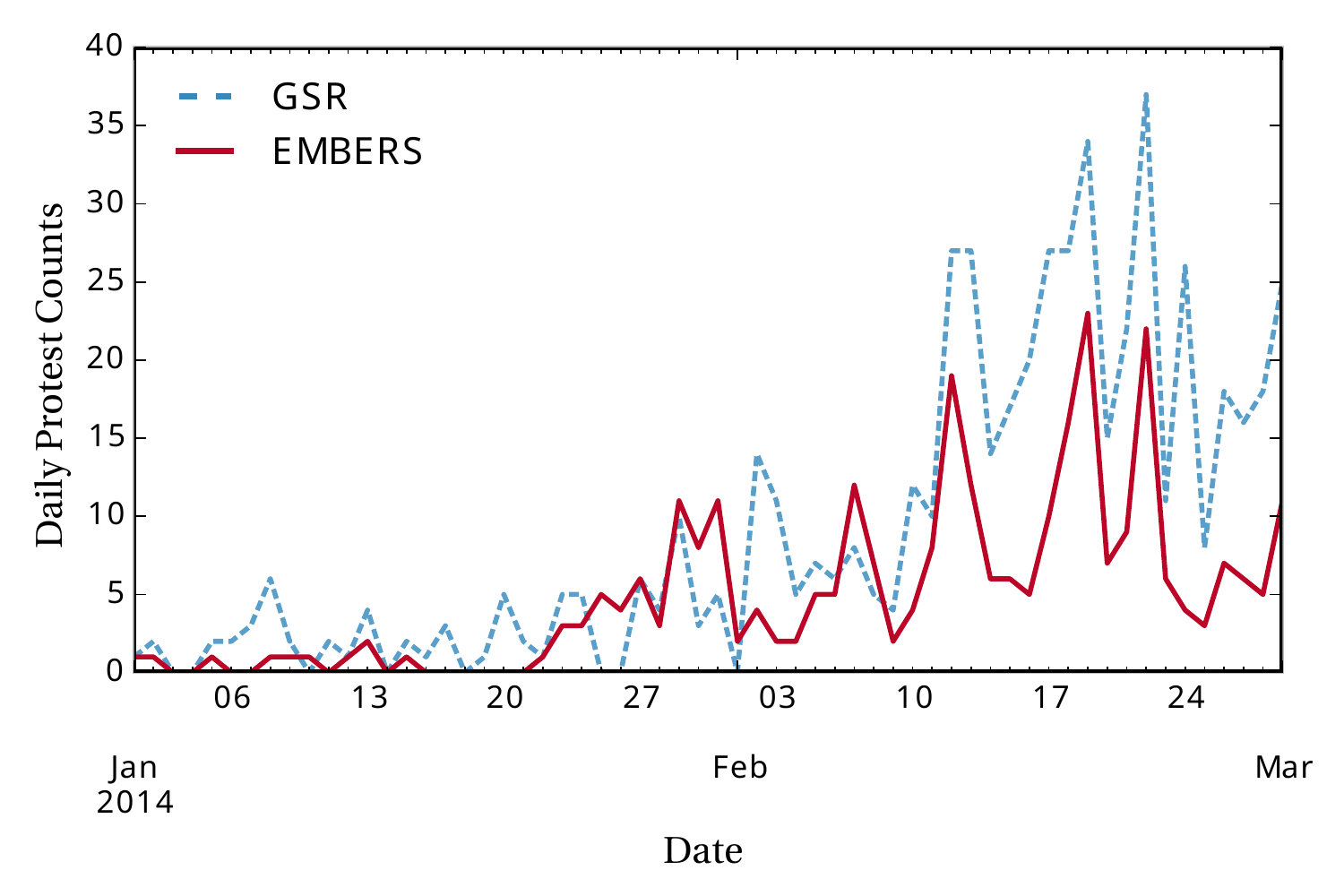}
\caption{EMBERS performance during the Venezuelan student protests (Feb-Mar 2014).}
\label{fig:venezuelaMarch14}
\end{figure}

Early 2014 Venezuela started experiencing a situation of turmoil with a
large portion of its population protesting due to insecurity, inflation
and shortage of basic goods. This period saw one of the highest levels of
civil disobedience in Venezuela with protests beginning in January with the murder
of a former Miss Venezuela. However, the protests started gaining more importance
 and turned violent and more frequent with students joining the movement following an
attempted rape of a student on campus in San Cristobal. EMBERS captured
some of these first `calls to protest` at San Cristobal and its nearby surrounding areas
 and correctly forecast the population (Education) and that the protests would turn violent.
A majority of the protesters were demanding that president Nicolas Maduro step down owing
to the poor economic policies and widespread corruption. EMBERS was capable of capturing
that the reason behind the protests were mainly against government policies with corruption being
a major theme.The EMBERS models working on twitter were also clearly able to identify some of major
leaders involved in the protest such as the major opposition leader Leopoldo Lopez.
Though the events mainly started off from San Cristobal it spread widely throughout the country, EMBERS
captured this spillover very well as shown in Figure~\ref{fig:venezuelaMap}.
Figure~\ref{fig:venezuelaMarch14}
shows how EMBERS closely forecast the spike in the number of events  during this period.

\subsubsection*{Mexico Protests (Oct 2014)}
\label{sec:mexico}
\begin{figure} \centering
\includegraphics[width=.8\columnwidth]{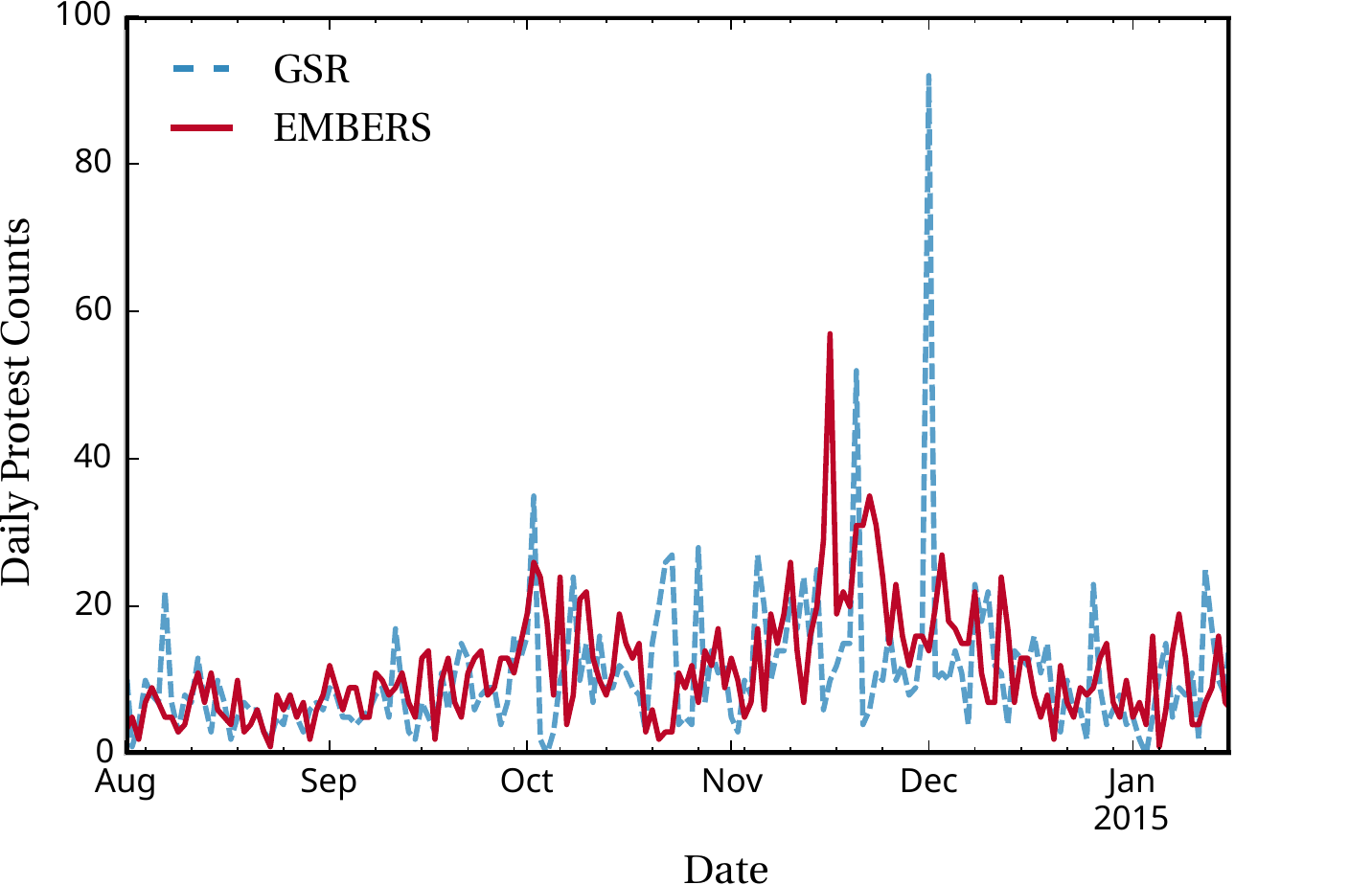}
\caption{EMBERS performance during Mexico protests (Oct 2014).}
\label{fig:mexicoOct14}
\end{figure}

\begin{figure*}
\centering
\includegraphics[width=.8\textwidth]{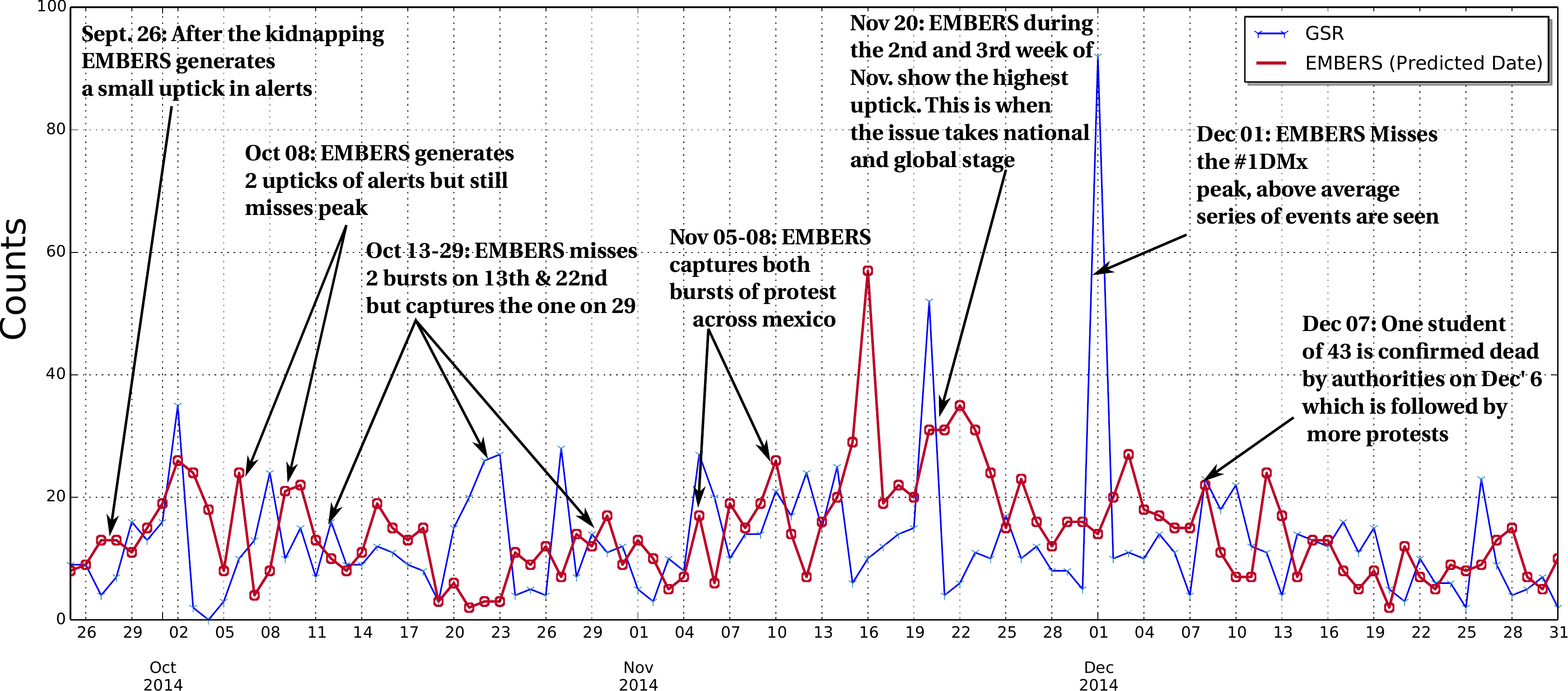}
\caption{Timeline of Mexico protests, showing the correspondence
between counts of GSR events and EMBERS alerts on a daily basis.}
\label{fig:mexicoTimeline}
\end{figure*}

Late September 2014, there were some peaceful protests by students from Ayotzinapa
in Mexico against discriminatory hiring practices for teachers. During
these protests,
police opened fire on the students killing around three and about 43 went missing. This poor
handling of the protest by the Mexican government caused widespread demonstrations throughout
the country over the next few months in support of the families of the 43 missing students.
A lot of these protests were violent in nature with demonstrators expressing extreme
dissatisfaction against the government and president Pena Nieto. EMBERS, as
shown in Figure~\ref{fig:mexicoOct14}
forecast an uptick in Mexico protests during early October 2014 with a lead time of about three days.
It also generated  a series of alert spikes coinciding with the first
large-scale nationwide protests between October 5th and 8th.
Figure~\ref{fig:mexicoTimeline} provides a timeline of GSR events and
EMBERS alerts for Mexico during this period. This figure provides a detailed
comparison of the continuous stream of alerts produced by EMBERS during this period with how
the actual events unfurled in the real world.

\subsubsection*{Colombia Protests (Dec 2014 to March 2015}
\begin{figure} \centering
\includegraphics[width=.8\columnwidth]{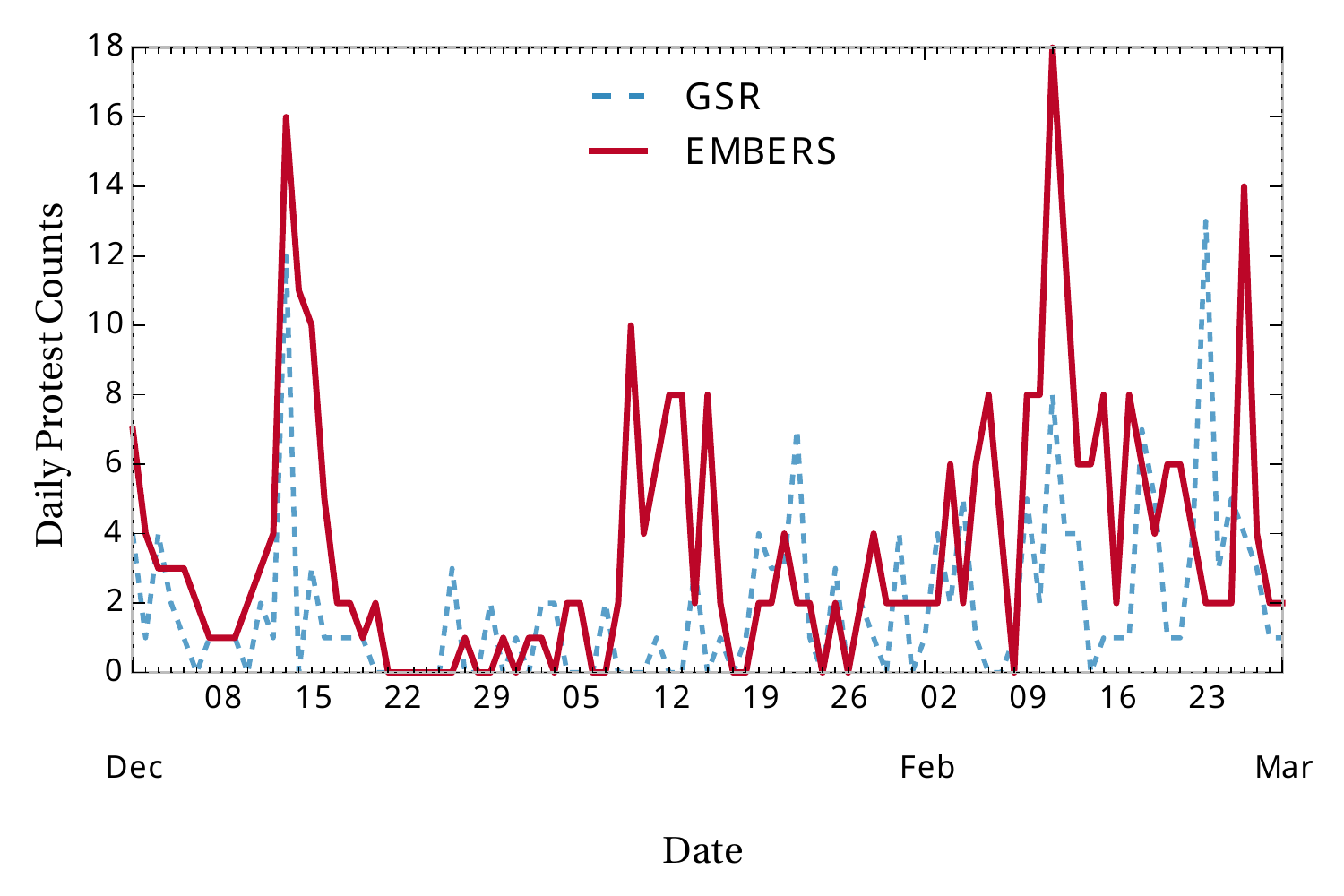}
\caption{Embers Performance during the Colombia Protests (Dec.2014 to Mar.2015)}
\label{fig:colombiaDec14}
\end{figure}

Colombia witnessed two different significant protests during this period, one during late December 2014
and the other during February 2015. Towards the end of 2014 the Colombian government were on the process
of moving forward with peace negotiations to end 50 years of conflict with the Revolutionary Armed Forces
of Colombia (FARC). With the FARC rebels having been associated with various acts of terror like extortion, armed conflict,
kidnapping, ransom, illegal mining etc., for a long period, the people of Colombia gathered in huge numbers
 to protest against possible amnesty for the FARC rebels.EMBERS successfully forecast the uptick in the number of events during the
middle of December 2014 as indicated in Figure~\ref{fig:colombiaDec14}. The figure also shows the increase in protest counts
during February 2015 though in this case EMBERS over-predicted the counts. The protests in February 2015
were mainly by led by truckers union demanding better freight rates, labor rights and against high fuel prices.
The truckers extended for about a month and caused huge losses of about \$300
million to the Colombian economy.
EMBERS picked up on the truckers protests right from the beginning but was wrong in over estimating the numbers during
February 11-12.

\subsubsection*{Paraguay Protests (February 2015)}
The February 2015 protests in Paraguay were mainly carried out by peasants
against the actions of the President Horacio Cartes.The protests were carried out after
president Horacio's public revelation that he had opened two private swiss bank accounts.
The protests also had a historical significance. It was also being carried out as a tribute
to peasant leaders and activists who were murdered. The peasants also protested
against the introduction of the new public-private partnership law.

\begin{figure} \centering
\includegraphics[width=.8\columnwidth]{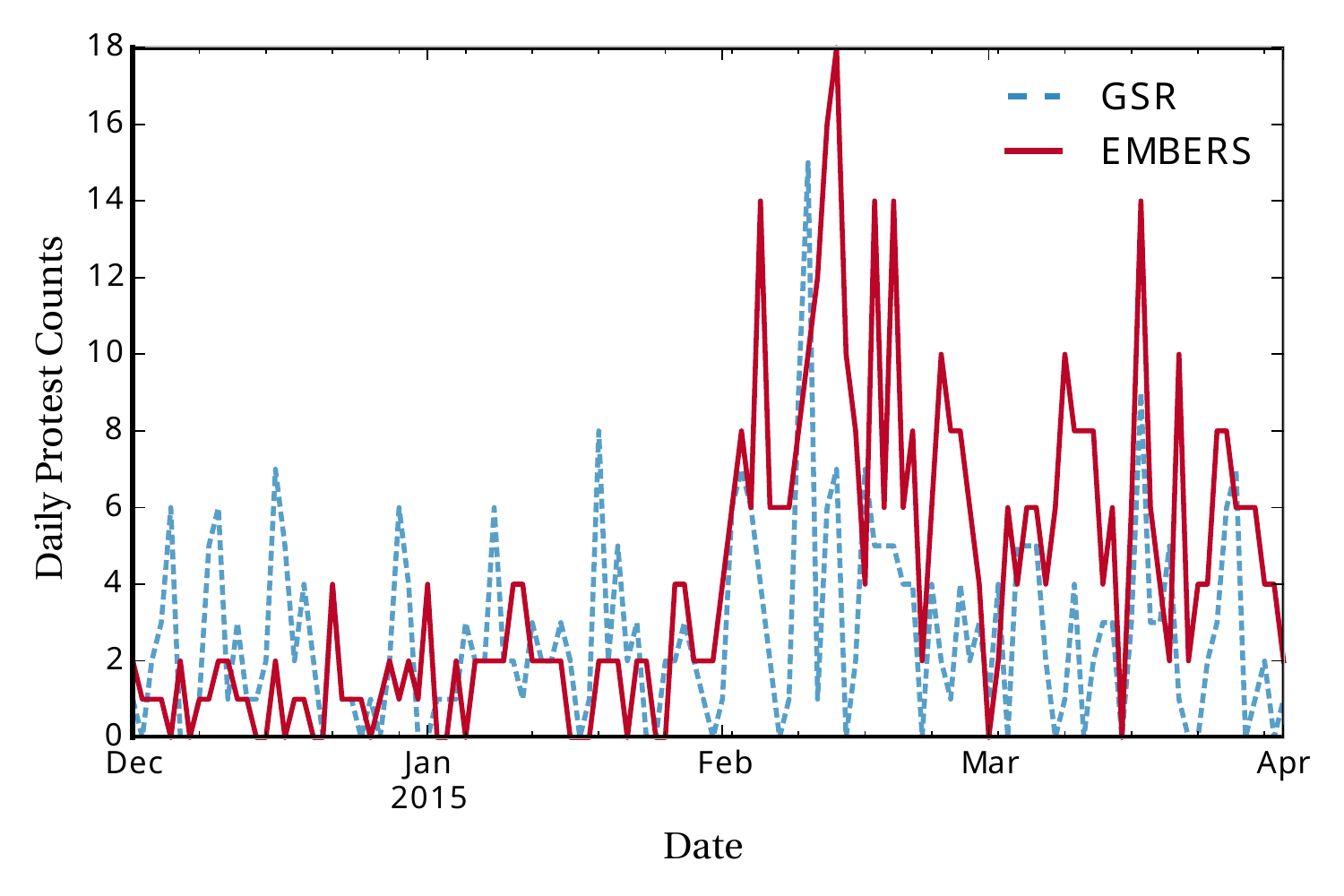}
\caption{EMBERS performance during Paraguay Protests (Feb. 2015)}
\label{fig:paraguay15}
\end{figure}

EMBERS forecast the uptick in number of protest events in Paraguay during mid
February 2015 as shown in Figure~\ref{fig:paraguay15}.

\subsection{EMBERS Misses}
Next, we outline specific large-scale events that
EMBERS failed to forecast accurately, along with a discussion of underlying
reasons.

\subsubsection{Brazilian Protests (March 2015)}
\begin{figure}
\centering
\includegraphics[width=.8\columnwidth]{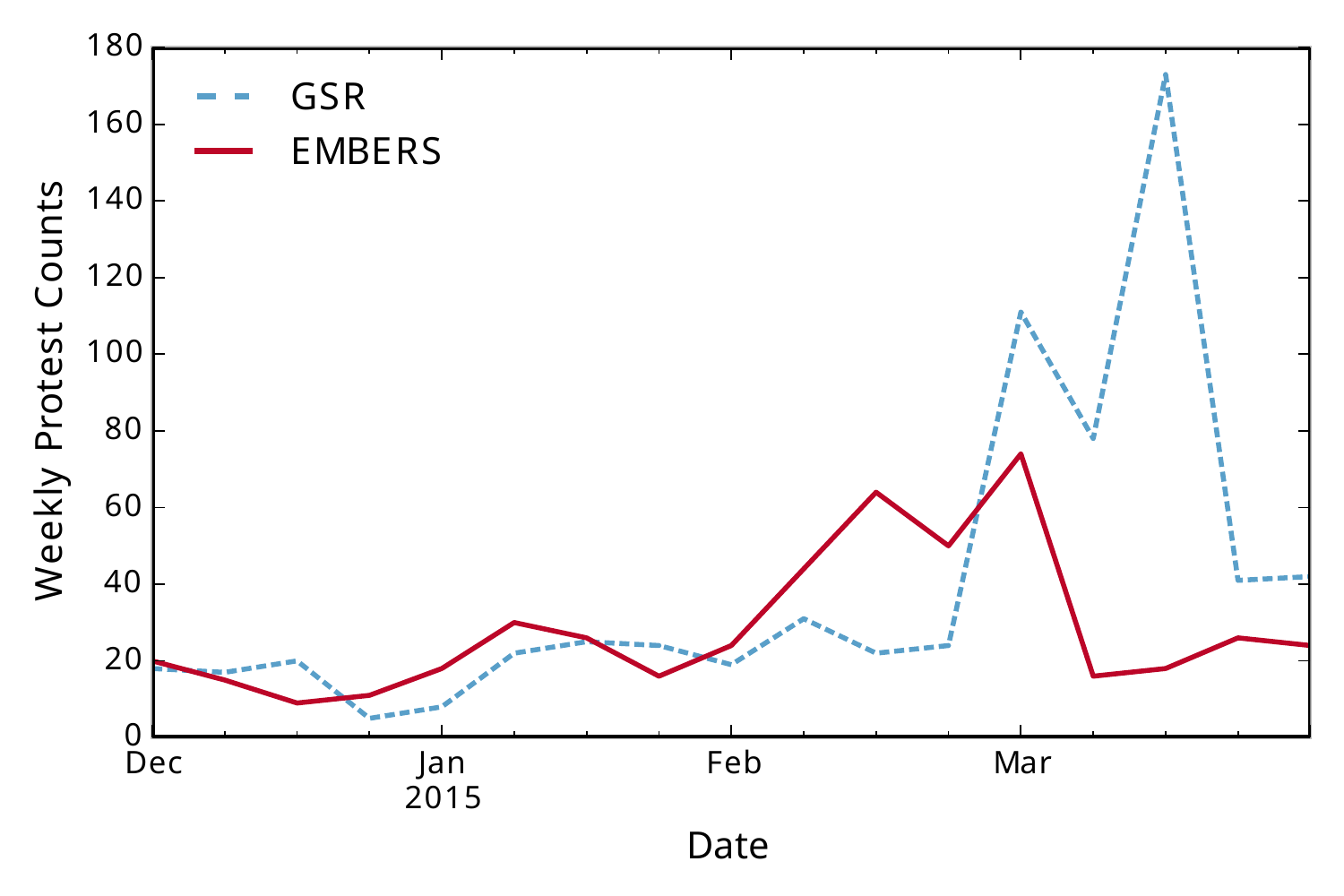}
\caption{Brazil 2015 Protests}
\label{fig:brazilSpring}
\end{figure}

The beginning of 2015 saw a series of protests in Brazil demanding the
removal of president Dilma Roussef amidst much furore against the increasing corruption in the country.
The protest magnitude increased significantly due to the revelations that many politicians belonging
to the ruling party accepted bribes from the state run energy company Petrobas. The protests saw huge participation from
the general population, with protesters generally estimated to be around a
million. EMBERS, as shown in Figure~\ref{fig:brazilSpring},
picked up the onset of events but failed to capture the sudden rise in the number of events.

During this period there was a significant architectural change in the
EMBERS processing pipeline. As mentioned in Section.~\ref{sec:background} the
EMBERS system enrichment pipeline consists of the following: natural language processing,
geocoding and relative time phrase normalization (temporal tagging). During early 2015
EMBERS had moved to the Heideltime~\cite{heideltime} temporal
tagger from the previously used TIMEN~\cite{timen} temporal tagger due to
Heideltime's support for more languages and an active development cycle as
opposed to TIMEN.
Heideltime had no support for Portuguese (the primary
language used in Brazil) and the EMBERS software development team had extended Heideltime to
support Portuguese by translating the underlying
resources for Spanish to Portuguese. As we learnt subsequently,
the simple translation of rules from Spanish to Portuguese
was not sufficient and this affected the recall of one of the key models
for Brazil, viz. the planned protest model.
Since the planned protest model relies
almost exclusively on the quality of information (specifically, date) extraction from text,
its performance significantly deteriorated. This was subsequently corrected for the future.

\subsubsection{\it Mexico Protests (Dec 2014)}
The December 2014 were a continuation of the series of protests that began in October 2014
as described in Section.~\ref{sec:mexico}. People turned out in huge numbers in different cities of Mexico demanding
President Pena Nieto's ouster owing to the manner in which the case of the 43 missing students were handled.
The protests were largely peaceful except for a few cases where vehicles were torched and windows and office equipment
were broken.

EMBERS missed the huge single day spike on December 1st. Though having predicted a nationwide event for
December 1, EMBERS failed to capture the individual cities where the protest were carried out and also was unable
to pick the magnitude. On manual analysis of the event, it was found that the date, December 1,  was picked by the protesters due to
its historical significance. This was the day when President Pena Nieto was sworn
in as President in 2012 amidst much controversy and opposition from
many specific constituent groups. The manual analysis also led to the understanding of how
special dates were mentioned by twitterati {\bf \#1Dmx}. Dates mentioned like this were totally unrecognised by the EMBERS
system and could have been one of the main reasons for EMBERS not being able to capture the peak on December 1st despite its
historical significance.

\subsubsection{Brazilian Spring Onset}
The EMBERS forecasts during the 2013 Brazilian Spring as shown in Fig.~\ref{fig:brazilJune13} was able to capture
the peak but as can be seen the system was unable to capture the initial
onset.  See Section~\ref{limitations} for a detailed discussion of
the limits of forecasting.

\section{Ablation Testing}
\label{sec:ablation}
\begin{figure}
\includegraphics[width=\columnwidth]{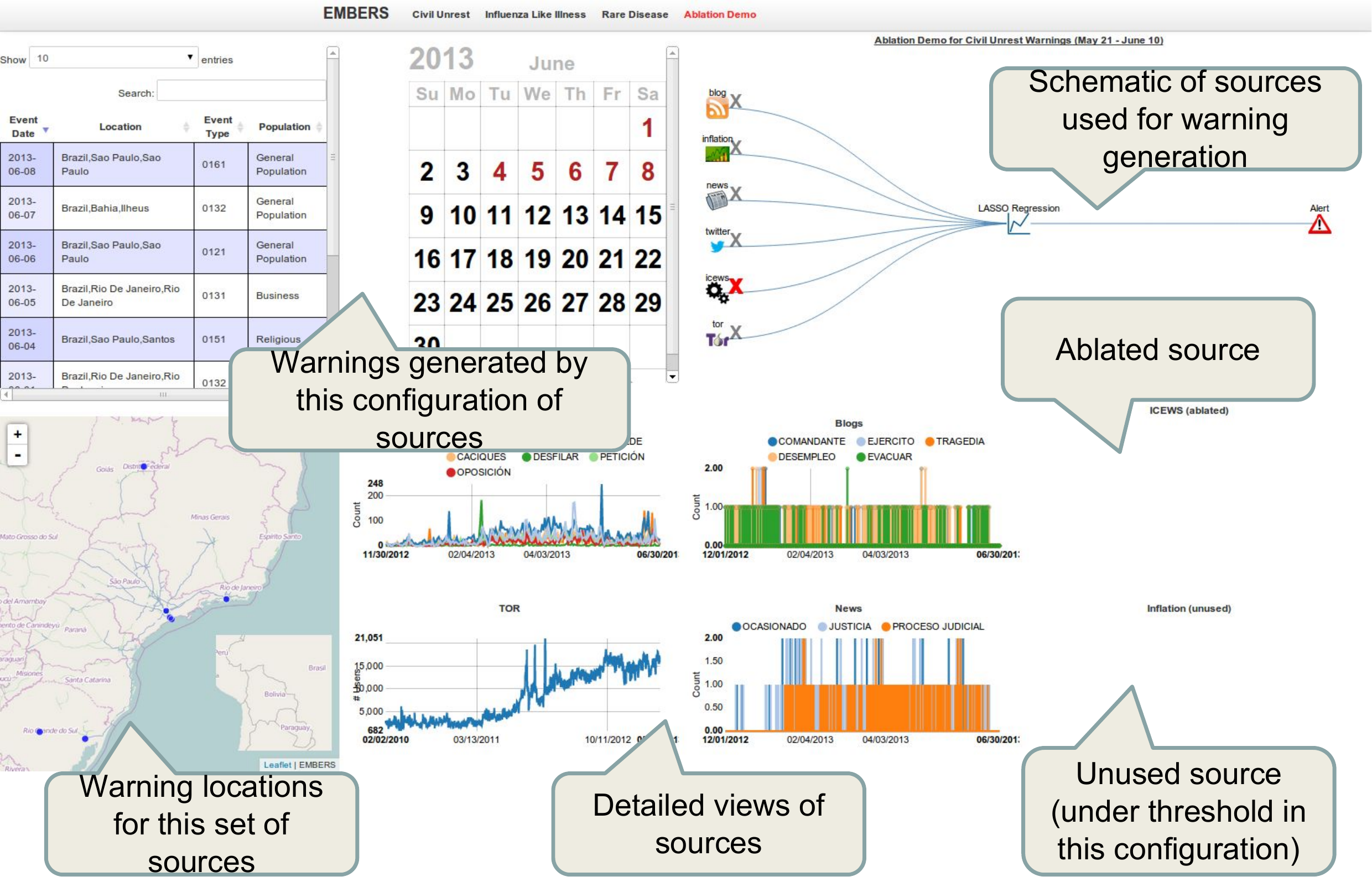}
\caption{EMBERS visualization on Ablation testing}
\label{fig:ablation}
\end{figure}

Different data sources provide different value to
the forecasting enterprise. It is important we understand the
value of a data source w.r.t. its forecasting potential.
In this section we describe ablation testing in EMBERS where the value added to forecasting performance by traditional media
sources like news and
blogs are compared to that provided by the social media sources.
Table~\ref{tb:ablation_twitter} shows the percentage difference in the performance metrics namely quality score, lead-time, precision and recall when only one
kind of source (traditional media or social media) is used with respect to the performance metrics when compared against the scenario when
both sources are used.
The table clearly shows that social media sources are mainly necessary in achieving high recall but not that useful in achieving high
lead times for which the traditional media sources are required. This
behavior is expected as social media is where daily chatter
occurs whereas signs of organization and calls for protest often
happen on traditional meadia.
Mainly, Table~\ref{tb:ablation_twitter} makes it evident that to build a successful forecasting system we need a good mix
of both traditional and social media sources.  Figure~\ref{fig:ablation} shows a snapshot of the EMBERS ablation visualizer. The visualizer provides an
analyst with the capability to selectively
remove data sources and assess the differences in final alerts.

\begin{table}
\small
\caption{Comparison of performance metrics with and without social media sources. Social Media sources contributes a lot towards recall but loses out on
lead-time.}
\label{tb:ablation_twitter}
\vspace{-3mm}
\resizebox{\columnwidth}{!}{
\begin{tabular}{|L|c|c|c|c|}
\hline
Data Source          & Quality-Score & Lead-time & Precision & Recall \\
\hline
Social Media Sources & -16.48\% &-55\% &+35\% &-14\% \\
\hline
Non-Social Media Sources & +8.42\% & +30\%  &+79\% &-33\% \\
\hline
\end{tabular}
}
\vspace{-5mm}
\end{table}

\section{Forecasting Surprising Events}
\label{sec:suprising}
The GSR contains a mix of everyday, mundane, protests as well as surprising events such as the Brazilian Spring.
We aimed to ascertain the relative ease of forecasting each class of events
with respect to a simple baserate model.

The baserate model generates alerts using the rate of occurrence of events in the past three months.

To define surprising events,
we employed a maximum entropy approach. For this purpose, each event is assumed
to be describable
in terms of three dimensions: country, population and event type. The GSR can then be conceptualized
as a cube. We infer a maximum entropy distribution conditioned on the marginals
induced by the cube using iterative proportional fitting~\cite{bishop2007discrete}, as shown in Algorithm.~\ref{algo:maxent}.
\begin{algorithm}
\caption{Surprise GSR calculation}
\begin{algorithmic}[1]

\Procedure{SURPRISE-GSR}{}
\State {\textbf{Input}: $\mathcal{G}= \{\mathcal{G}_{-1},\mathcal{G}_{-2}, \mathcal{G}_{-3}\}$}
\State {\textbf{Output}: Maximum likelihood estimate for <event-type, population, country> tuple -- $\mathcal{\hat{M}}$}

\State Each event in the GSR-$\mathcal{G}$ is mapped to a three dimensional vector of <event-type, population, country>. Each such vector
       is mapped to a cell in a 3-D cube and the value $x_{ijk}$ for a cell in the cube represents the frequency of the vector <i, j, k>.
       $m_{ijk}$ represents the maximum likelihood estimate for a given cell.

\State Initialize $m_{i,j,k}=1 \forall i,j,k$
\For {$c \in \{0,MAX\_ITERATIONS\}$}
\For {$i,j,k \forall \textsc{<Event-Type, Population, Country>}$}
\State $\hat{m}^{c+1} = {\hat{m}^c}_{ijk}*\frac{x_{ij+}}{{\hat{m}^c}_{ij+}} $

\State $\hat{m}^{c+2} = {\hat{m}^{c+1}}_{ijk}*\frac{x_{i+k}}{{\hat{m}^c}_{i+k}} $

\State $\hat{m}^{c+3} = {\hat{m}^{c+2}}_{ijk}*\frac{x_{+jk}}{{\hat{m}^c}_{+jk}} $

\EndFor

\If {$\hat{m}^c - \hat{m}^{c-1} < \tau$}

\Return {$\hat{m}^c$}

\EndIf

\EndFor

\Return {$\hat{m}^c$}

\EndProcedure

\end{algorithmic}

\label{algo:maxent}
\end{algorithm}

Every month, events from the last three months of the
GSR is used to populate the underlying cube of counts and the iterative proportional fitting procedure in Algorithm.~\ref{algo:maxent} is
used to estimate the maxent values for each cell. The resultant sample counts are then scaled to
match the observed number of GSR events in the current month. The cell-wise difference between the inferred
maxent distribution and the observed GSR is computed and all cells with significance greater than five standard
deviations are classified as containing surprising events. In essence, this approach takes the GSR as input and
creates a truncated GSR against which we can evaluate EMBERS (and the baserate model).

\begin{table}
\caption{Surprising events inferred by maxent analysis.}
\renewcommand{\arraystretch}{1.1}
\vspace{-3mm}
 \centering
 \begin{tabular}{|l|l|m{5cm}|}
 \hline
May '13  &  Colombia	 &  Nationwide protests against administrative policy changes regarding the
Youth in Action program and social security pension \\ \hline
Jun '13  &  Brazil  &  Brazilian Spring \\ \hline
Jul '13  &  Brazil  &  Brazilian Spring \\ \hline
Sep '13  &  Mexico  &  Nationwide protests against education and energy reforms \\ \hline
Oct '13  &  Mexico  &  Nationwide protests against education and energy reforms \\ \hline
Oct '13  &  Uruguay  &  Nationwide protests demanding increase in minimum wage \\ \hline
Feb '14  &  Venezuela	  &  Venezuelan Student Protests \\ \hline
Mar '14  &  Venezuela  &  	Venezuelan Student Protests \\ \hline
May '14  &  Brazil  &  Nationwide demonstrations in response to the 2014 FIFA World Cup and other social issues \\ \hline
Jun '14  &  Brazil  &  Nationwide demonstrations in response to the 2014 FIFA World Cup and other social issues \\ \hline
Aug '14  &  Argentina	  &  Nationwide protests against drops in wages, employment, and rising inflation \\ \hline
Sep '14  &  Ecuador  &  Nationwide protests to demand changes in labor policies \\ \hline
Oct '14  &  Mexico  &  Nationwide protests after the discovery of mass graves of kidnapped students  \\ \hline
\end{tabular}
\label{tab:maxentEvents}
\end{table}

\begin{figure} \centering
\includegraphics[width=.99\columnwidth]{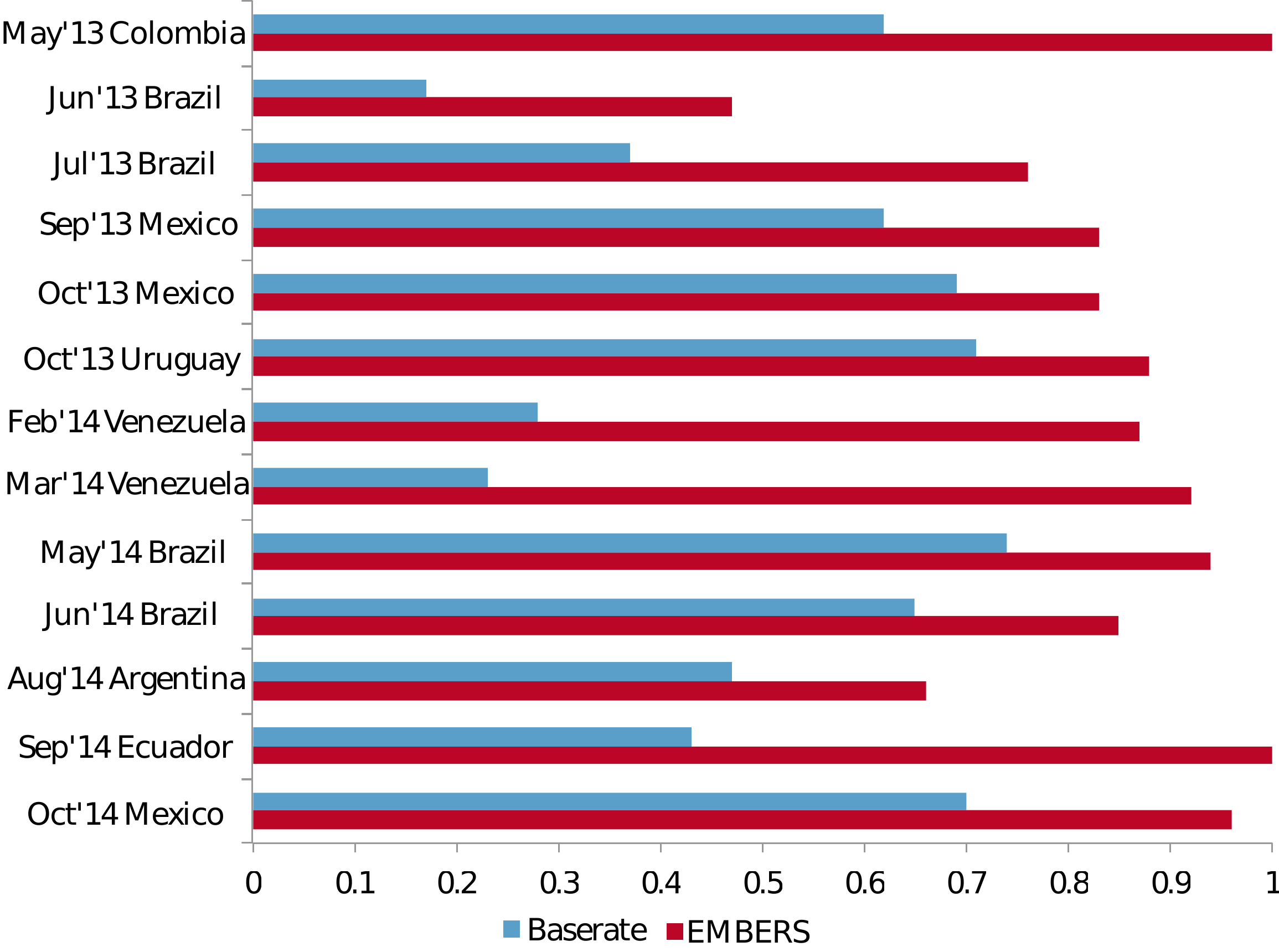}
\caption{Performance of EMBERS vs a baserate model for surprising events.}
\label{fig:maxent}
\end{figure}

In Table~\ref{tab:maxentEvents} we present several inferred events in Latin America from our maximum entropy
filter.
For all these events, we compare the recall of both EMBERS and baserate model
in Figure~\ref{fig:maxent}.
It is clear that EMBERS is able to forecast these significant upticks consistently.

\section{Uncertainties in Forecasting}
\label{limitations}
\begin{figure}
\includegraphics[width=\columnwidth]{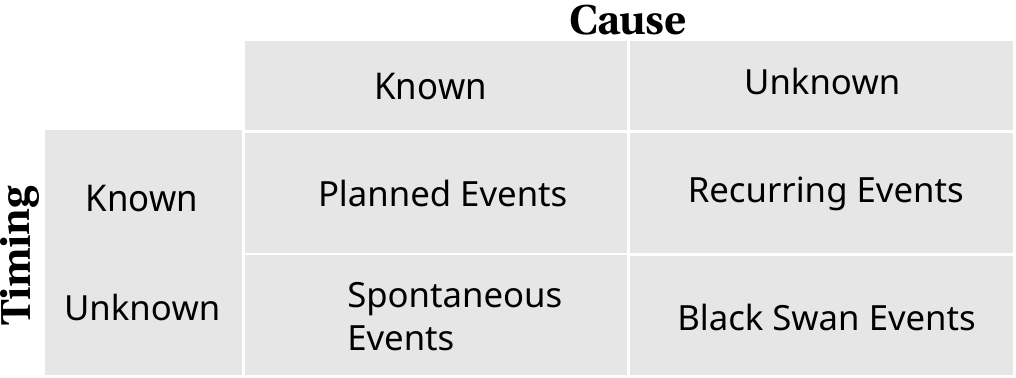}
\caption{Studying the limitations of event forecasting.}
\label{rumsfeld}
\end{figure}

While examining the events of civil unrest closely
in the past few years, it was clear to our
team that events carry two distinct types of uncertainties: {\bf cause}
and {\bf timing}.
Fig.~\ref{rumsfeld} summarizes these uncertainties.

Among all the incidents of civil unrest that we encounter, the largest and the most significant ones
are planned events.  These events are usually organized by political parties, labor and student unions.
Since it takes a huge effort to organize protest demonstrations that
attract thousands, the organizers must disseminate information regarding the venue
and the date and time.  These announcements are posted on the organizers’ websites and
are widely shared on social media.  By scouring our sources,
it has been possible for EMBERS to accurately forecast the occurrence of these types of protests.

The recurring events take place on a regular basis.  For instance, in Chile and Argentina the
`mothers of the disappeared' protest the disappearance of their children by the military dictatorships
of the 1970's and 19780's on a certain particular day of the week and in the same plaza.
In some countries with large Muslim populations, fighting and protests break out regularly after
Friday evening prayer as people stream out of the mosques after listening
to fiery sermons. These are typically small events but if they are reported as part of our GSR,
EMBERS models will be able to forecast them.

The protests for which the causes are known but not the timing are staged spontaneously.  These
events are the outcomes of longstanding frustration and anger which fuel widespread
protests in response to trigger events.  Thus, the viral videos of police brutality or
a sudden change in government policy can start a prairie fire of protests.  The Brazilian
Spring with origins in bus fares and which channeled
public anger against corruption and government mismanagement is a classical example. The challenge here
is not just to be aware of the underlying tensions that might erupt when an event occurs, but to also
distinguish between events that do and do not perform as triggers. Algorithms to better model
precursors is an area of further research that will aid further forecasting this class of events.

Finally, {\it black swan} events~\cite{taleb-book} are rare and truly unforeseen and can happen as a
result of natural disasters, the sudden death of a leader, or even the sudden rise of a
small group that can truly destabilize a nation.  For instance the rise of the
Islamic State in Iraq and Syria (ISIS) has truly confounded policymakers all over the world.
While there were other Sunni groups, from al-Qaeda to al-Nusra, that contributed to
instability, the rapid ascendance of ISIS, which did not depend on an isolated terrorist attack and
burst out with a clear holding of territories as a full-scale insurgency, surprised most observers.
It might not be feasible to forecast the beginnings of such events; however, once such movements
have been initiated, models should be able to detect and forecast their momentum.
\section{Ethical Issues}
\label{sec:ethics}
EMBERS, as an anticipatory intelligence system, has many powerful legitimate uses but is also
susceptible to abuse.

First, it is important to have a discussion of civil unrest and its role in society.
In the proper circumstances,
civil unrest enhances the ability of citizens to communicate not only their views but also
their priorities to those who govern them. Governments constantly need to make choices and find
it difficult to know, on specific issues at particular times, how their constituencies value the available
options. Elections are retrospective indicators and rarely issue-specific; polling taps into sentiment,
but is not a good indicator of priorities or strength of feeling because of the
low cost associated with responding. Events, on the other hand, indicate a willingness to
bear some costs (organization, mobilization, identification) in support of an issue and
thus reveal not only preferences but provide some indication of priorities.

An open sources indicators approach, as used here, is a potentially powerful
tool for understanding the social construction of meaning and its translation into behavior.
EMBERS can contribute to making the transmission of citizen preferences to
government less costly to the economy and society as well. There are economic
costs to even peaceful disruptions embodied in civil unrest due to lost work
hours and the deployment of police to manage traffic and the interactions
between protestors and bystanders. Given the vulnerability of large gatherings
to provocation by handfuls of violence-oriented protestors (e.g., Black Box
anarchists in Brazil) the economic, social and political costs of
large-scale public demonstrations are also potentially significant to marchers, bystanders, property owners
and the government -- democratically elected or not. The right to demonstrate can still
be respected but if the government responds to grievances in time,
the protestors may cancel the event or fewer people might participate in the event.
In today's interconnected society, protests also cause disruptions to supply chain logistics, travel, and
other sectors, and anticipating disruptions is key to ensuring safety as well as reliability.

The potential power of civil unrest forecasting systems, like those
of most scientific advances, is susceptible to abuse by
both democratic and non-democratic governments. The appropriate safeguards require
developing transparent and accountable democratic systems, not outlawing science.
Non-democratic governments may clearly abuse such forecasting systems. But even
here the value of forecasting civil unrest is not simply negative. Many non-democratic regimes transition to democratic
ones, often in a violent process but not always (in Latin America, authoritarian regimes negotiated
transitions to democracy without a civil war in Mexico, Honduras, Peru, Bolivia, Brazil,
Uruguay, Argentina and Chile). The rational choice models of authoritarian decision-making in
such crises always explain a dictatorships’ collapse rather than accommodation to
a transition by pointing to the lack of credible information in
a dictatorship regarding citizens’ true feelings. EMBERS-like models may thus provide the
information that facilitates and encourages some transitions.

\section{Conclusions}
We have presented the successes as well as the lessons learned from the EMBERS
architecture as a result of four years of 24x7 operation.  EMBERS has shown
itself to be a reliable predictor of civil unrest events in 10 different
countries, and three different major languages.  Yet despite these successes
EMBERS has also been a learning experience and even in its misses much has been
learned.

Future work falls primarily in two directions..
First, we will continue
to build a strong capability to forecasting societal events,
leveraging multiple, overlapping,
models of selective superiority.
Being able to better understand how multiple
alerts from different models should be fused would 
be an invaluable tool to analysts.
Secondly, we are investigating
techniques to remove or reduce the human element required in generating 
the GSR.  Currently the most human intensive part of the EMBERS project 
is generating a GSR for training and validation of the models.

\end{document}